\documentclass[conference]{IEEEtran}

\usepackage[hidelinks]{hyperref}
\usepackage{cite}
\usepackage{amsmath,amssymb,amsfonts}
\usepackage{rotating,multirow,multicol}
\usepackage{booktabs}
\usepackage{multirow}
\usepackage{array}
\usepackage{algorithmic}
\usepackage{graphicx}
\usepackage{textcomp}
\usepackage{xcolor}
\usepackage{float}
\usepackage{subfig}
\usepackage{pifont}
\usepackage{url}
\usepackage{balance}

\begin{document}

\title{The IJCNN 2025 Review Process}

\author{\IEEEauthorblockN{Michele~Scarpiniti, Danilo~Comminiello}\\
\IEEEauthorblockA{\textit{Department of Information Engineering, Electronics and Telecommunications}\\
\textit{Sapienza University of Rome}\\
via Eudossiana 18, 00184 Rome, Italy\\
\{michele.scarpiniti,danilo.comminiello\}@uniroma1.it}
}

\maketitle

\begin{abstract}
The International Joint Conference on Neural Networks (IJCNN) is the premier international conference in the area of neural networks theory, analysis, and applications. The 2025 edition of the conference comprised 5,526 paper submissions, 7,877 active reviewers, 426 area chairs, 2,152 accepted papers, and more than 2,300 attendees. This represents a growth of about 100\% in terms of submissions, 200\% in terms of reviewers, and over 50\% in terms of attendees as compared to the previous edition. In this paper, we describe several key aspects of the whole review process, including a strategy for ranking the scores provided by the reviewers by evaluating a score index and a calibrated version used experimentally to remove reviewer-specific bias from reviews.
\end{abstract}


\section{Introduction}
\label{sec:intro}

The IJCNN 2025 review process was a massive undertaking, with 12,407 authors from 68 countries submitting 5,526 papers. A team of 426 area chairs (ACs) and 7,877 active reviewers contributed a total of 18,996 reviews (with an average of 3.44 reviews per paper). Ensuring fairness at such a scale posed a significant challenge. To manage this complexity, the process needed to be as decentralized as possible at every stage.

The role of the Technical Program Chairs (TPCs) was very difficult, since each stage of the process showed its peculiar difficulties and challenges \cite{Shah2022}. These steps concerned the choice and setting of the conference management system, finding and inviting reviewers and meta-reviewers, interacting with authors and sending all notifications and communications, and, overall, preparing the review/meta-review forms and making decisions.

As the conference management system, we chose CMT\footnote{See: \url{https://cmt3.research.microsoft.com/About}}, thanks to its several advantages for managing large academic conferences, including its scalability, security, and customizable features.

Due to the large number of papers, the review process has been organized in a hierarchical manner: three TPCs coordinated a set of 425 ACs, each of whom handled the entire review process of about 15/20 papers. At least 4 reviewers have been assigned to each paper. The initial reviewers were assigned by TPCs, but ACs were able to change and/or add new reviewers if needed, in case of they were not responding. This hierarchy allowed to complete the review process in time, with only the 2.2\% of papers with only two reviews, while all the other papers received three or more reviews (with an average of 3.44 reviews per paper).

As the conference does not have a rebuttal period, based on the received review reports a score index has been evaluated, in order to provide a ranking of all the papers. Then the top ranked papers have been accepted according to a fixed acceptance rate that should not exceed the 40\%. As well known, quantifying judgment is a very difficult task \cite{Merkhofer1987}, due to the subjectivity of the review process: different reviewers can judge differently the same paper. To this aim, also an experimental procedure to calibrate reviews by removing possible reviewer-specific biases has been implemented. This procedure has already been used in other famous conferences, such as NeurIPS 2014 and 2016 \cite{Cortes2021,Shah2018}.

Although an exciting and stimulating experience, the management of large conferences is not free of huge problems, issues, and challenges \cite{Shah2022,Cortes2021}, which need to be handled by the Chairs. These embrace both technical aspects, as the configuration of the conference system and providing support to a large number of authors, that decision making, as founding the best strategy to obtain a fair paper ranking. We remark that handling such issues is very time consuming.

In the rest of the paper, we briefly describe the general review process in Section~\ref{sec:general} and the reviewers assignment in Section~\ref{sec:assignment}. Section~\ref{sec:decision} introduces the used procedure to rank papers and take the final decisions, while Section~\ref{sec:calibration} discusses about an experimental approach to remove reviewer-specific biases. A discussion is provided in Section~\ref{sec:discussion}. Finally, Section~\ref{sec:conclusion} concludes the paper.

\section{The general process}
\label{sec:general}

The whole process of managing a large conference is very challenging and involves a huge number of different steps \cite{Shah2022}. Chairs need to take into account for many choices during this process \cite{Shah2018}. Along with traditional steps, some conferences have also implemented some experimental approaches to render the decision process more fair, trying to calibrate the reviewers' scores. A good example of this conference is the Neural Information Processing Systems (NIPS), in particular the edition 2014 and 2016 \cite{Shah2018}.

In the first meeting, TPCs discussed about the general setup and organization of the entire process, by proposing a hierarchical organization of the review process composed of three key roles: Chair, Area Chair (which is called Meta-Reviewer in CMT), and Reviewer. Chairs have also composed the submission, review, and meta-review forms, and configured the CMT system for the submission phase and the subsequent review phase.

The first challenge was the recruitment of a sufficient number of reviewers. Reviewers were retrieved from three sources: 1) some lists of past conferences, included IJCNN 2023; 2) the principal contact authors of the submissions; and, 3) volunteer reviewers, who provided their availability by filling a Google form available on the conference website. Reviewers of the first group were recruited at the beginning of the submission phase, those of the third group were retrieved during the whole submission process, while the second group was recruited after the submission phase was closed. This third group of reviewers was necessary due to the unprecedented large number of submissions received. In CMT, we use some user tags to identify these reviewers: Rev1 (reviewers of the first group), Rev2 (principal contact authors), Rev3 (expert volunteers), and Rev4 (student volunteers). Table~\ref{tab:reviewers} summarizes the number of reviewers that accepted the invitation among the 13,766 invited. Then 7,736 reviewers, about the 85\%, have effectively performed at least one review. Of these, 7,525 reviewers worked on the main track, related to regular papers, special session papers, and the special track on ``Human-AI interaction in creative arts and sciences''.

\begin{table}[htb]
    \centering
    \caption{Reviewer distribution at IJCNN 2025.}
    \label{tab:reviewers}
    \begin{tabular}{lcrr}
    \toprule
       \textbf{Group} & \textbf{Tag} & \textbf{Number} & \textbf{Completed} \\
    \midrule
       Group 1  & Rev 1 &  3,702 &  3,174 \\
       Group 2  & Rev 2 &  4,064 &  3,351 \\
       Group 3  & Rev 3 &  1,272 &  1,059 \\
       Group 3  & Rev 4 &    164 &    152 \\
    \midrule
       Total    &       &  9,202 &  7,736 \\
    \bottomrule
    \end{tabular}
\end{table}

A distribution of the reviewers among the different conference tracks and the number of received review reports per track are summarized in Table~\ref{tab:reports}. Some reviewers, 141 for sake of precision, have been used in different tracks, increasing virtually the number of reviewers to 7,877. An average 2.41 papers per reviewer has been assigned.

\begin{table}[htb]
    \centering
    \caption{Reviewers per track and related received reports.}
    \label{tab:reports}
    \begin{tabular}{lrr}
    \toprule
       \textbf{Track}   & \textbf{Number of Reviewers} & \textbf{Reports} \\
    \midrule
       Main             &  7,527 & 18,427 \\  
       Competition      &     10 &     17 \\ 
       Position Papers  &     73 &    130 \\  
       Workshops        &    267 &    432 \\ 
    \midrule
       Total            &  7,877 & 18,996 \\ 
    \bottomrule
    \end{tabular}
\end{table}

The selection of Area Chairs (ACs) was only on volunteer basis. ACs of the regular papers were recruited among people which provided their availability by filling a Google form available on the conference website, while ACs of the other tracks (special sessions, competitions, workshops, etc.) are the organizers of the track itself. A summary of the ACs is shown in Table~\ref{tab:metareviewers}.

\begin{table}[htb]
    \centering
    \caption{Area Chair (AC) distribution at IJCNN 2025.}
    \label{tab:metareviewers}
    \begin{tabular}{llr}
    \toprule
       \textbf{Track}     & \textbf{Tag} & \textbf{Number}  \\
    \midrule
       Regular            & RegAC  &  262  \\
       Special Sessions   & SpeSAC &  120  \\
       Special Track      & STrAC  &    2  \\
       Competitions       & CompAC &    4  \\
       Position Papers    & PosPAC &    1  \\
       J2C Papers         & J2CAC  &    1  \\
       Workshops          & WorkAC &   36  \\
    \midrule
       Total              &        &  426 \\
    \bottomrule
    \end{tabular}
\end{table}

The submission received at IJCNN 2025 were more than 5,500, distributed among tracks as shown in Table~\ref{tab:submissions}. This posed a serious challenge in handling the whole review process. All these submissions have been assigned to at least 4 reviewers in order to guarantee a minimum number of reviews. With the help of ACs, we were able to receive 18,996 reviews, which correspond to an average of 18,996/5,526 $\approx$ 3.44 reviews per paper. An excellent result at this scale!

Along with papers published in the conference proceedings, IJCNN 2025 has also considered submissions for other four tracks with non-archival papers. Specifically, both Competition and Workshop tracks considered for short papers, i.e., 2 to 4 pages papers about ideas, already published works, or preliminary results, which will be presented during the conference as a normal paper but without the publication on the proceedings. INNS 2025 also welcomes the submission of Journal-to-Conference (J2C) papers, i.e., the presentation at IJCNN 2025 of recently published and high-impact journal papers on topics related to methodological advancements and compelling novel applications of neural networks and deep learning systems. Finally, a dedicated Poster Track has been organized to present preliminary results as a poster presentation during the conference as a normal paper but without the publication on the proceedings. Details about submission and acceptance for these non-archival track are shown in the bottom part of Table~\ref{tab:submissions}.

\begin{table}[htb]
    \centering
    \caption{Submission distribution at IJCNN 2025.}
    \label{tab:submissions}
    \begin{tabular}{lrrr}
        \toprule
    \textbf{Track} & \textbf{Submissions} & \textbf{Accepted}  \\
    \midrule
       Regular                    & 4,225 & 1,533  \\
       Special Sessions           & 1,049 &   478  \\
       Special Track              &    30 &    12  \\
       Competitions               &    11 &     7  \\
       Position Papers            &    66 &    15  \\
       Workshops                  &   145 &   107  \\
    \midrule       Total          & 5,526 & 2,152  \\
    \midrule
       Competition Short papers   &     1 &     1  \\
       Workshop Short papers      &    30 &    23  \\
       J2C Papers                 &    33 &    27  \\
       Poster Track               &    58 &    58  \\
    \midrule
       Total                      &   122 &   109  \\
    \bottomrule
    \end{tabular}
\end{table}

As the IJCNN conference does not consider a rebuttal phase with a corresponding paper revision, the next step was to evaluate a ranking of all the submitted papers (as described in the following Section \ref{sec:decision}), in order to select the top one according to a maximum acceptance rate \cite{Douceur2009}. This year, INNS board asked us to not exceed the 40\% of acceptance rate to be competitive in conference international rankings. The proposed ranking of the papers has also been constructed by taking into consideration some experimental procedures, as removing the reviewer-specific bias from the reviews.

Details of the accepted paper per track are summarized in Table~\ref{tab:submissions}, from which the acceptance rate was: $2,152/5,526 = 38.94$\%. Hence, we have satisfied the constraint of 40\% by also taking a small margin. At the same time, all papers have been checked for plagiarism by using the iThenticate service, directly accessible on the CMT platform.

\section{Reviewers assignment}
\label{sec:assignment}

The reviewer assignment is a challenging phase, not only for the large number of reviewers to handle, but also to avoid any misconduct \cite{Shah2022}. For example, a first problem is the so-called \emph{lone wolf}: since some reviewers are also authors, the reviewer could increase the chances of acceptance of their own papers by manipulating the reviews of other papers, such as providing lower ratings. A second type of misconduct is a potential collusion between reviewers and authors; this risk is potential high since we asked reviewers to provide a bid for the papers. In addition, also matching the reviewers' expertise with the papers is difficult in general. Several approaches have been proposed in literature \cite{Shah2022,Stelmakh2021}, however most of them try to  maximize the total sum of the similarity scores of all assigned reviewer-paper pairs.

In order to mitigate these potentials risks of misconduct, we take some actions. First of all, at most one (author) reviewer (in group Rev 2 of Table~\ref{tab:reviewers}) is assigned to each paper so to mitigate the effect of lone wolf. Second, during the assignment phase, we use some weight on three different criteria exploited to associate papers to reviewers. The reviewers' assignment exploits three different criteria: reviewers' bids, Subject Area relevance, and TPMS scores. The Toronto Paper Matching System (TPMS) is a service to match papers with reviewers, developed by researchers at the University of Toronto \cite{Charlin2013}. To mitigate the potential collusion risk, the reviewers' bids have been assigned a low weight (20\%), while a 30\% is reserved for the TPMS (very useful but many reviewers did not register to the service), and a 50\% for the Subject Area relevance, which is the easiest and most powerful criterion. CMT simply performs a convex combination of these three data sources.

To better mathematically formalize the assignment process, let $\mathcal{P}$ be a set of $P$ papers indexed by $p \in \mathcal{P}$, $\mathcal{R}$ a set of $R$ reviewers indexed by $r \in \mathcal{R}$, and $\mathcal{M}$ a set of $M$ meta-reviewers indexed by $m \in \mathcal{M}$. We denote with $\mathcal{A}_r$ the set of assigned reviewer-paper pairs $\mathcal{A}_r = \left\{ (r, p) \in \mathcal{R} \times \mathcal{P} \right\}$, and with $\mathcal{A}_m$ the set of assigned meta-reviewer-paper pairs $\mathcal{A}_m = \left\{ (m, p) \in \mathcal{M} \times \mathcal{P} \right\}$. For every paper $p$ assigned to reviewer $r$ and a meta-reviewer $m$, the reviewer provides a quantized score $s_{rp} \in \mathbb{Q}$, while the meta-reviewer provides a quantized score $\xi_{mp} \in \mathbb{Z}$, both bounded by specific intervals.

\section{Making decision}
\label{sec:decision}

\subsection{Review scores}
\label{sec:review_score}

In the review form, reviewers are asked to answer to a number of specific questions related to the following aspects:
\begin{enumerate}
    \item Reviewer's confidence ($R_c$)
    \item Four quality criteria:
    \begin{itemize}
        \item Relevance to IJCNN ($C_1$)
        \item Technical quality ($C_2$)
        \item Novelty ($C_3$)
        \item Quality of presentation ($C_4$)
    \end{itemize}
    \item Award quality ($AQ$)
    \item Overall recommendation ($S$)
    \item Comments for authors
\end{enumerate}
Except for ``Comments for authors'', which is a textual field containing detailed comments for authors and suggestions to improve the paper, all the other indicators are numeric, assuming values in specific sets. Specifically, the weights assigned to the quality criteria $C_1$--$C_4$ and to the Reviewer's confidence $R_c$ are shown in Table~\ref{tab:criteria_weights}, while the weights assigned to the overall recommendation $S$ provided by reviewers and meta-reviewers are shown in Table~\ref{tab:recommendation_weights}, which is mapped on a symmetric 7 point Likert scale \cite{Likert1932}. For $S$ and $C_1$--$C_4$ criteria we opted for a symmetric score. The weight assigned to the award quality $AQ$ is simply the set $\{0, \, 1\}$, where 1 denotes an award nomination.

\begin{table}[htb]
    \centering
    \caption{Weights assigned to the quality criteria $C_1$--$C_4$ and the Reviewer's confidence $R_c$.}
    \label{tab:criteria_weights}
    \setlength{\tabcolsep}{8pt}
    \begin{tabular}{lrr}
    \toprule
    \textbf{Option} & $\mathbf{C_1}$--$\mathbf{C_4}$ & $\mathbf{R_c}$  \\
    \midrule
        Excellent     & $~1.0$   & 1.2  \\
        Very good     & $~0.5$   & 1.1  \\
        Good          & $~0.0$   & 1.0  \\
        Fair          & $-0.5$   & 0.9  \\
        Poor          & $-1.0$   & 0.8  \\
    \bottomrule
    \end{tabular}
\end{table}

\begin{table}[htb]
    \centering
    \caption{Weights assigned to the overall recommendation provided by reviewers and meta-reviewers.}
    \label{tab:recommendation_weights}
    \setlength{\tabcolsep}{10pt}
    \begin{tabular}{lr}
    \toprule
    \textbf{Option} & $\mathbf{S}$ \\
    \midrule
        Strong Accept  & $~3$  \\
        Accept         & $~2$  \\
        Weak Accept    & $~1$  \\
        Borderline     & $~0$  \\
        Weak Reject    & $-1$  \\
        Reject         & $-2$  \\
        Strong Reject  & $-3$  \\
    \bottomrule
    \end{tabular}
\end{table}

Given an assignment set $\mathcal{A}_r$, each $r$-th reviewer contributes to the $p$-th paper with a \textit{score} $s_{rp}$ evaluated as follows:
\begin{equation}
s_{rp} = S_{rp} + \sum_{k=1}^4 C_{k,rp} + AQ_{rp} ,
\label{eq:rev_score}
\end{equation}
where the subscript $rp$ denote the score and a specific metric for the $p$-th paper assigned by the $r$-th reviewer, and $C_k$, $k=1,2,3,4$, are the four quality criteria.

Let $R$ be the number of reviewers for the $p$-th paper. The aggregated score $s_p$ for the $p$-th paper is obtained by the weighted mean of the scores with the reviewers' confidences:
\begin{equation}
s_p = \frac{ \sum_{r=1}^R R_{c,rp} \, s_{rp}}{\sum_{r=1}^R R_{c,rp}}.
\label{eq:paper_rev_score}
\end{equation}
All these per-paper scores are packed into the following score vector $\mathbf{S}_R$:
\begin{equation}
\mathbf{S}_R = \left[ s_1, s_2, s_3, \ldots, s_P \right]^T,
\label{eq:score_rev_vector}
\end{equation}
where $P$ is the total number of reviewed papers. The vector $\mathbf{S}_R$ of reviewers' scores of each paper is then normalized by mapping the values inside the interval $[0, \, 100]$ to obtain the proposed \textit{normalized reviewers' score} $\widetilde{\mathbf{S}}_R$:
\begin{equation}
\widetilde{\mathbf{S}}_R = 100 \times \frac{\mathbf{S}_R - \mathbf{S}_{R\min}}{\mathbf{S}_{R\max} - \mathbf{S}_{R\min}},
\label{eq:score_rev_normalized}
\end{equation}
where $\mathbf{S}_{R\min}$ and $\mathbf{S}_{R\max}$ are the minimum and maximum value in the vector $\mathbf{S}_R$, respectively.

\subsection{Meta-Review scores}
\label{sec:meta-review_score}

Moving to consider the meta-reviewer score, this is fundamentally based on the overall recommendation $S$ with the same weights in Table~\ref{tab:recommendation_weights}. We can highlight that usually a single meta-reviewer is assigned to a paper. Hence, the aggregated score $\xi_p$ for the $p$-th paper coincides with the score $\xi_{mp}$, given an assignment set $\mathcal{A}_m$. However, we have some few cases with more than one Meta-Reviewer, for example in some special sessions where more than an organizer provided independent scores. In these cases, we use the mean values of scores $\xi_{mp}$ for the different meta-reviewers $m$. Similar to \eqref{eq:score_rev_vector}, these scores are stacked into a vector $\mathbf{S}_M$:
\begin{equation}
\mathbf{S}_M = \left[ \xi_1, \xi_2, \xi_3, \ldots, \xi_P \right]^T.
\label{eq:score_metarev_vector}
\end{equation}
Also the vector $\mathbf{S}_M$ of meta-reviewers' scores of each paper is then normalized by mapping the values inside the interval $[0, \, 100]$ to obtain the proposed \textit{normalized meta-reviewers' score} $\widetilde{\mathbf{S}}_M$ applying an equation similar to \eqref{eq:score_rev_normalized}.

\subsection{Paper scores}
\label{sec:paper_score}

Having evaluated the (averaged) reviewers' scores $\widetilde{\mathbf{S}}_R$ and the meta-reviewer's scores $\widetilde{\mathbf{S}}_M$ for each paper, we have now to find a consensus between these two scores by providing a final per paper \textit{score index} $\mathbf{SI}$. Different approaches are possible to meet a consensus between reviewers and meta-reviewer(s). Just to refer few approaches, a simple mean can work effectively. However, also more sophisticated approaches are possible, such as algebraic methods based on the Dempster's rule of combination \cite{Shafer2016,Haenni2008} or fuzzy approaches \cite{Liu2012}. However, the choice of the membership functions in this last approach makes this method too subjective.

We decided that a good approach is to make the final score as a number in between the reviewers and meta-reviewer ones, mitigating the different points of view that reviewers and meta-reviewer may have. Instead of using a simple mean computation, we opted for using the harmonic mean between the two scores \cite{Wilson2019}:
\begin{equation}
\mathbf{SI} = \frac{2 \times \widetilde{\mathbf{S}}_R \times \widetilde{\mathbf{S}}_M}{\widetilde{\mathbf{S}}_R + \widetilde{\mathbf{S}}_M},
\label{eq:harmonic_mean}
\end{equation}
where the multiplication and division are computed element-wise. Papers are then sorted in descending order according to their score index $\mathbf{SI}$ and the top $a$\% of papers are selected to meet the acceptance rate $a$ of the conference. For IJCNN 2025, we set $a = 40$\%.

\section{Experimental approach}
\label{sec:experimental}

In order to have a different point of view, some experimental approaches have been implemented \cite{Ge2015}. Specifically, we have implemented the score dequantization proposed in \cite{Liu2022} and the review calibration introduced in \cite{Cortes2021} and \cite{Shah2018}.

\subsection{Review dequantization}
\label{sec:dequantization}

The review process usually requires a quantized score, i.e. reviewers are asked to provide scores from a constant number of quantization levels \cite{Liu2022}. The gap between consecutive quantized scores can generate inaccuracies in the evaluation of a paper and can result in a loss of information, thereby contributing to the difficulty in making decisions regarding the acceptance of papers. To minimize this negative quantization effect, we try to dequantize the received score by using the approach proposed in \cite{Liu2022}.

To this aim, we need to minimize the following cost function:
\begin{equation}
\mathcal{L}_q = \sum_{(r,p) \in \mathcal{A}_r} \left( \widehat{s}_{rp} - \widehat{\overline{s}}_p \right)^2 + \lambda \sum_{(r,p) \in \mathcal{A}_r} \left( \widehat{s}_{rp} - s_{rp} \right)^2.
\label{eq:dq_costFunction}
\end{equation}
where $\widehat{s}_{rp}$ is the dequantized score and $\widehat{\overline{s}}_p$ is the average of the dequantized reviewers' score for paper $p$.

Hence, we are interested in solving the following minimization problem:
\begin{equation}
\begin{split}
& \min_{\widehat{s}_{rp}} \mathcal{L}_q \\
& s.t. ~ s_{rp} - 0.25 \leq \widehat{s}_{rp} \leq s_{rp} + 0.25,
\end{split}
\end{equation}
where the constraint is due to the fact that the reviewer score $s_{rp}$ has a quantization of 0.5 based on Eq. \eqref{eq:rev_score} and weights in Tables \ref{tab:criteria_weights} and \ref{tab:recommendation_weights}. The solution of the previous optimization problem lead to the following result \cite{Liu2022}:
\[
\widetilde{y} = \frac{1 + n_p \lambda}{n_p (1 + \lambda)} s_{rp} + \sum_{(r',p)\in\mathcal{A}_r, r'\neq r} \frac{1}{n_p (1 + \lambda)} s_{r'p}
\]
\begin{equation}
\widehat{s}_{rp} = \begin{cases}
    \widetilde{y}, & \text{if} ~ \widetilde{y} \in \left[ s_{rp}-0.25, \, s_{rp}+0.25 \right] \\
    s_{rp} - 0.25, & \text{if} ~ \widetilde{y} < s_{rp} - 0.25 \\
    s_{rp} + 0.25, & \text{if} ~ \widetilde{y} > s_{rp} + 0.25 \\
\end{cases}
\label{eq:dequantized_score}
\end{equation}
In previous equations, $n_p$ is the number of reviews for the $p$-th paper, while $\lambda > 0$ is an hyper-parameter. For large $\lambda$ we have $\widehat{s}_{rp} \rightarrow s_{rp}$, while for $\lambda = 0$ we have $\widehat{s}_{rp} = s_{rp} \pm 0.25$. 

Since, generally, only one meta-reviewer is assigned, there is no need to provide a meta-reviewer dequantization procedure.

\subsection{Review calibration}
\label{sec:calibration}

The main problem of assessing the quality of a paper (but it is a more general problem when a set of objects has to be evaluated by a panel of assessors), is to handle reviewers' miscalibrations \cite{MacKay2017}. In fact, a score of 6 out of 10 is different for different reviewers \cite{Wang2019}. This because the reviewers' rating scales may be miscalibrated since it is subjective and not universal. The main problem is due to the fact that this is an unsupervised problem, since we do not know the correct score to be assigned to a paper.

This is an old problem \cite{DeGroot1974} and several methodologies have been proposed in literature to handle such miscalibrations. Specifically, some approaches attempt an algebraic aggregation of judgments \cite{Yager1988} or the application of the so called \textit{opinion calculus} \cite{Haenni2008}, which proposes a particular application of the Dempster-Shafer theory fusing opinions by means of the opinion triangles. There also exist some least squares (LS) approaches \cite{Tan2021,Roos2011}, or the approach in \cite{Roos2011} based on quadratic programming. However, these methods are too computationally complex due to the large number of reviews to analyze. A more statistical approach has been presented by Platt and Burges in \cite{Platt2012}, based on a regularized LS model, which is similar to a two-way ANOVA model, and was used in many NIPS conferences between 2002 and 2012. 

In 2013, Zoubin Ghahramani and Max Welling used a Bayesian extension of this model \cite{Ge2015}, while, more recently, outside the NeurIPS community, MacKay et al. in \cite{MacKay2017} have proposed a Bayesian approach that takes confidence scores into account. Like Welling and Ghahramani, also Cortes and Lawrence in \cite{Cortes2021} used a Bayesian variant of the Platt-Burges model, but formulating it as a Gaussian process. This approach has also been used in next NIPS conferences \cite{Shah2018}. For IJCNN 2025 we decided to follow fundamentally the approach introduced by Cortes and Lawrence in \cite{Cortes2021}.

We suppose that the review $r$ score $s_{rp}$ for paper $p$ can be decomposed as the summation of three terms \cite{Cortes2021}:
\begin{equation}
s_{rp} = q_p + b_r + \varepsilon_{rp},
\label{eq:calibration_model}
\end{equation}
where $q_p$ is the objective quality of paper $p$, $b_r$ is a reviewer-specific bias term taking into account for its personal judgment way (i.e., different reviewers interpret the scale differently), and $\varepsilon_{rp}$ is a subjective estimate of the quality of paper $p$ according to reviewer $r$. This implies that we need to estimate $P + R + PR$ values and may be computational burden for large conferences as IJCNN 2025 (for which this number is of the order of $43.5 \times 10^6$). However, in practice, since the system is sparse we do not need to estimate any parameters where no assignment is done for a pair paper-reviewer.

We can assume that the subjective quality of the paper $\varepsilon_{rp}$ is drawn from a Gaussian distribution with zero mean and variance $\sigma_{\varepsilon}^2$:
\[
\varepsilon_{rp} \sim N(0, \sigma_{\varepsilon}^2).
\]
We also assume that the objective quality $q_p$ is normally distributed with mean $\mu_q$ and variance $\sigma_q^2$:
\[
q_p \sim N(\mu_q, \sigma_q^2).
\]
Finally, we assume that the reviewer bias $b_r$ is a zero mean normally distributed latent variable with variance $\sigma_b^2$:
\[
b_r \sim N(0, \sigma_b^2).
\]
Hence, we need to estimate the four parameters: $\mu_q$, $\sigma_{\varepsilon}^2$, $\sigma_q^2$, and $\sigma_b^2$.

From \eqref{eq:calibration_model} we have that the reviewer $r$ score to paper $p$, $s_{rp}$, is a combination of three Gaussian-distributed factors. Hence, the \emph{a priori} marginal variance of the score of a reviewer-paper assignment is the sum of the three components: $\sigma_{\varepsilon}^2$, $\sigma_q^2$, and $\sigma_b^2$. Instead, the mean is constant and provided only by the quality $\mu_q$, which can be therefore estimated as the mean of the subjective quality $s_{rp}$. Cross-correlations between reviewer-paper assignments occur if either the reviewer is the same (when the cross-covariance is given by $\sigma_b^2$) or the paper is the same (when the cross-covariance is given by $\sigma_q^2$) \cite{Cortes2021}. Hence, the considered joint model for reviewer scores is as follows:
\begin{equation}
\mathbf{s} \sim N(\boldsymbol{\mu}_q, \mathbf{K}),
\label{eq:calibration_model_distribution}
\end{equation}
where $\mathbf{s}$ is a vector of stacked scores $s_{rp}$, $\boldsymbol{\mu}_q$ is the vector of the constant means $\mu_q$ and the elements of the covariance matrix $\mathbf{K}$ are given by:
\begin{equation}
k(i,j;k,l) = \delta_{i,k} \sigma_q^2 + \delta_{j,l} \sigma_b^2 + \delta_{i,k} \delta_{j,l} \sigma_{\varepsilon}^2,
\label{eq:calibration_covarianceMatrix}
\end{equation}
where $i$ and $j$ are the index of the paper and reviewer in the rows of $\mathbf{K}$ and $k$ and $l$ are the index of the paper and reviewer in the columns of $\mathbf{K}$.

Cortes and Lawrence \cite{Cortes2021} suggest to rewrite \eqref{eq:calibration_covarianceMatrix} as:
\begin{equation}
\begin{split}
k(i,j;k,l) &= \sigma_q^2 \left( \delta_{i,k} + \delta_{j,l} \frac{\sigma_b^2}{\sigma_q^2} + \delta_{i,k} \delta_{j,l} \frac{\sigma_{\varepsilon}^2}{\sigma_q^2} \right) \\
           &= \sigma_q^2 \left( \delta_{i,k} + \delta_{j,l} \widehat{\sigma_b}^2 + \delta_{i,k} \delta_{j,l} \widehat{\sigma_{\varepsilon}}^2 \right).
\end{split}
\label{eq:calibration_covarianceMatrix_2}
\end{equation}
This simplifies the problem since $\sigma_q^2$ can be optimized with a fixed-point approach, by leaving only two free parameters that can be explored on a grid \cite{Cortes2021}. Since the mean value $\mu_q$ is estimated as the mean of the subjective quality $s_{rp}$, the multivariate probability density function of $\mathbf{s}$ in \eqref{eq:calibration_model_distribution} can be explicitly written as:
\begin{equation}
p_{\mathbf{s}}(\mathbf{s}) = \frac{1}{\sqrt{(2\pi)^R \left|\mathbf{K}\right|}} e^{- \frac{1}{2}\mathbf{s}^T \mathbf{K}^{-1} \mathbf{s}},
\label{eq:calibration_model_distribution2}
\end{equation}
where $R$ is the number of reviewers. Given the scaling done in \eqref{eq:calibration_covarianceMatrix_2}, we can set $\widehat{\mathbf{K}} = \mathbf{K}/\sigma_q^2$ and computing its negative log-likelihood:
\begin{equation}
\mathcal{L} = -\log p_{\mathbf{s}}(\mathbf{s}) = \frac{R}{2} \log\!\left(2\pi \sigma_q^2\right) + \frac{1}{2} \log \left|\widehat{\mathbf{K}}\right| + \frac{1}{2 \sigma_q^2} \mathbf{s}^T \mathbf{K}^{-1} \mathbf{s}.
\label{eq:negative_loglikelihood}
\end{equation}

\begin{figure*}[htb]
    \centering
    \subfloat{\includegraphics[width=0.455\linewidth]{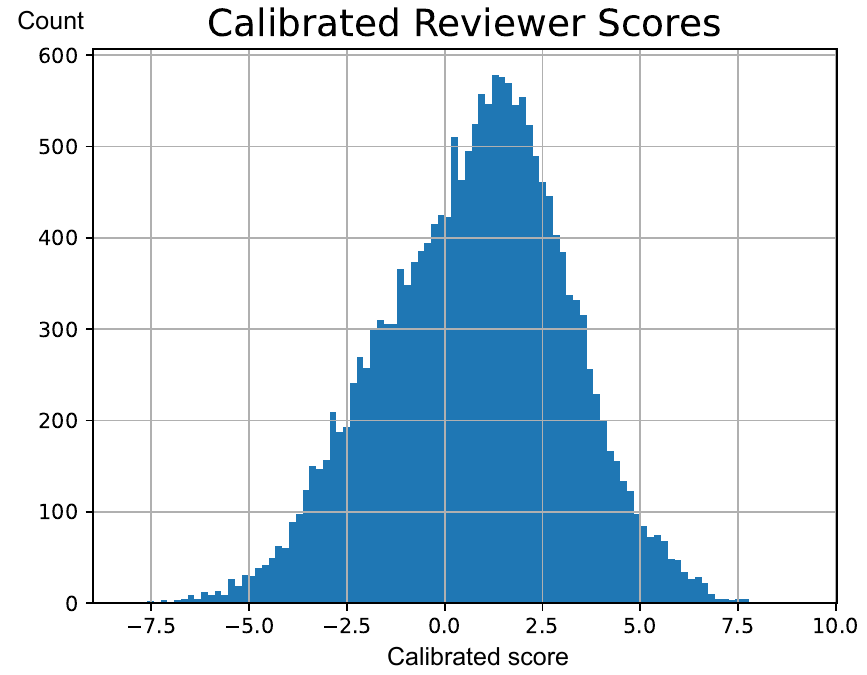}\label{fig:Rev_cal_hist}}
    \hspace{1em}
    \subfloat{\includegraphics[width=0.455\linewidth]{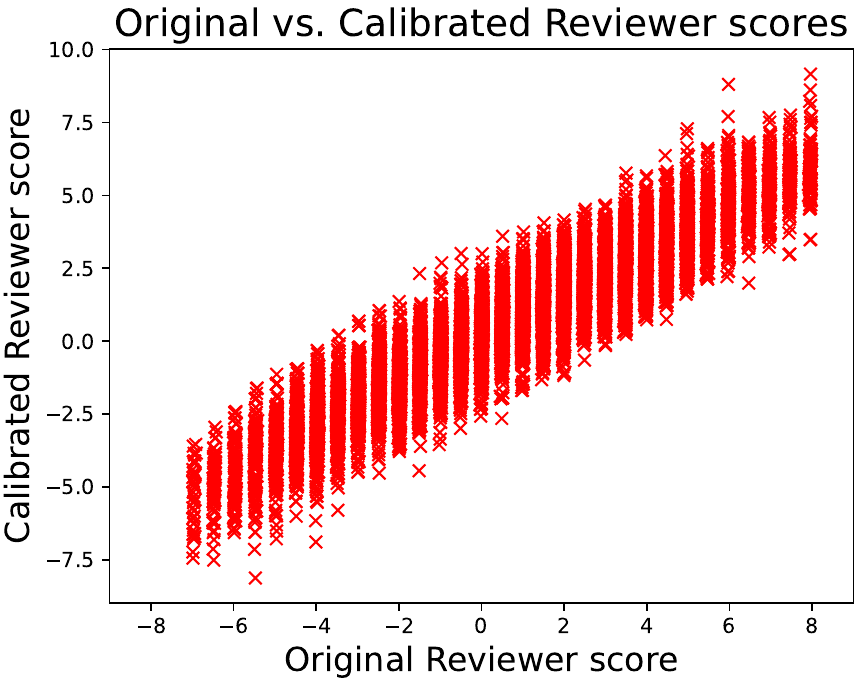}\label{fig:Rev_cal_scatter}}
    \caption{Calibrate reviewers' scores: (a) histogram of the calibrated scores $\mathbf{y}$ in \eqref{eq:conditional_distribution}, (b) scatterplot of the original vs. calibrated reviewers' scores.}
    \label{fig:calibrated_scores}
\end{figure*}

\noindent Hence the optimal value for the $\sigma_q^2$ parameter is evaluated by setting to zero the negative log-likelihood in \eqref{eq:negative_loglikelihood}, obtaining:
\begin{equation}
\sigma_q^2 = \frac{1}{R} \, \mathbf{s}^T \mathbf{K}^{-1} \mathbf{s}.
\end{equation}
After that these parameters are evaluated by the maximum likelihood approach, we can remove the offset from the model by:
\begin{equation}
y_{rp} = q_p + \varepsilon_{rp},
\label{eq:calibration_model_output}
\end{equation}
This calibrated variable has a covariance matrix $\mathbf{K}_y = \mathbf{K}_q + \sigma_{\varepsilon}^2 \mathbf{I}$, and a cross covariance between $\mathbf{s}$ and $\mathbf{y}$ is also given by $\mathbf{K}_y$. Hence, the conditional distribution over the paper scores with the bias removed can be found as \cite{Cortes2021}:
\begin{equation}
\mathbf{y} | \mathbf{s}, \sigma_q^2, \sigma_b^2, \sigma_{\varepsilon}^2 \sim N(\boldsymbol{\mu}_y, \boldsymbol{\Sigma}_y),
\label{eq:conditional_distribution}
\end{equation}
where: 
\[
\boldsymbol{\mu}_y = \mathbf{K}_y \mathbf{K}^{-1} \mathbf{s} \qquad \text{and} \qquad \boldsymbol{\Sigma}_y = \mathbf{K}_y - \mathbf{K}^{-1} \mathbf{K}_y \mathbf{K}_y.
\]
We can now use $\boldsymbol{\mu}_y$ as the calibrated quality score. Figure~\ref{fig:calibrated_scores} shows the histogram of the calibrated scores $\mathbf{y}$ and the scatterplot between the original and calibrated scores.

\begin{figure}[htb]
    \centering
    \includegraphics[width=0.91\linewidth]{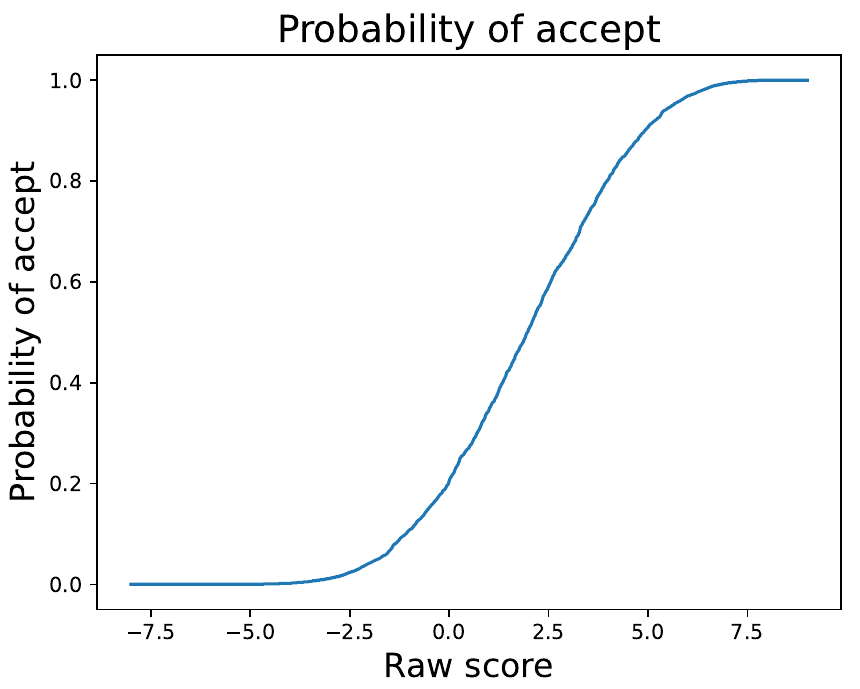}
    \caption{The per-paper acceptance probability with respect to the calibrated score provided by reviewers.}
    \label{fig:acceptance_probability}
\end{figure}

This posterior distribution of bias-adjusted scores can be used to estimate a per-paper acceptance probability. By sampling by this distribution, we can obtain a sampled conference. If we do that several times (e.g., 1,000 times), we can see how many times each paper was accepted to get a probability of acceptance. This acceptance probability with respect to the calibrated score is depicted in Figure~\ref{fig:acceptance_probability}.

\subsection{Aggregating reviews and meta-reviews}
\label{sec:aggregation}

\begin{figure}[htb]
    \centering
    \includegraphics[width=0.91\linewidth]{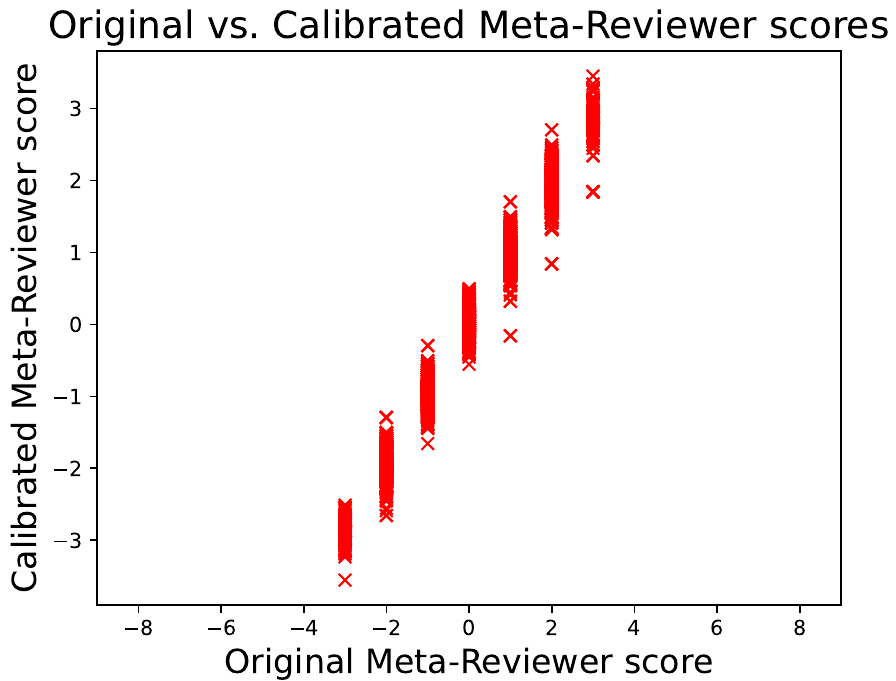}
    \caption{The scatterplot of the original vs. calibrated meta-reviewers' scores.}
    \label{fig:AC_cal_scatter}
\end{figure}

The calibrated scores $\boldsymbol{\mu}_y$ are then normalized in the $[0, \, 100]$ interval as done in \eqref{eq:score_rev_normalized}, by obtaining a calibrated version of the normalized reviewers' score $\widetilde{\mathbf{S}}_{R, cal}$. A similar calibration procedure can be adopted for the meta-reviewer scores $\xi_{mp}$ by obtaining another set of normalized and calibrated meta reviewers' scores $\widetilde{\mathbf{S}}_{M, cal}$. The scatterplot between the original and calibrated scores of meta-reviewers are shown in Figure~\ref{fig:AC_cal_scatter}.

Again, as done in Section~\ref{sec:paper_score}, the calibrated and normalized scores of reviewers $\widetilde{\mathbf{S}}_{R, cal}$ and meta-reviewers $\widetilde{\mathbf{S}}_{M, cal}$ can be merged together by using the harmonic mean in \eqref{eq:harmonic_mean} to generate the calibrated score index $\mathbf{SI}_{cal}$, which can be used by its comparison with respect to the non-calibrated version in \eqref{eq:harmonic_mean}.

\section{Discussion}
\label{sec:discussion}

The experimental procedure used to produce the calibrated score index $\mathbf{SI}_{cal}$ helped the final decision on the acceptance of papers in a gray area. Specifically, we look papers where both the rankings ($\mathbf{SI}$ and $\mathbf{SI}_{cal}$) agree in positive or negative and we have directly accepted and rejected these papers. Instead, where the rankings disagree an additional investigation on papers has been done by carefully reading the Meta-reviewers' suggestion and Reviewers' reports.

Grossly, this procedure implied that all papers which received an average score in between ``Strong Reject'' and ``Borderline'' in Table~\ref{tab:recommendation_weights} were rejected, while papers that received an average score  of ``Strong Accept'' and ``Accept'' were accepted. The gray area concerned fundamentally many papers that received an average score of ``Weak Accept''. Unfortunately, to meet the constraint on the acceptance rate, many of these papers have bee rejected.

As the huge number of challenges in handling large conferences, there is room for many improvements. Just to mention few points, as future works we can include LLM models to automatically analyze both reviews and papers to detect text generated by AI tools; in addition, it could be also useful to construct a metric based on the cosine similarity between all the received reviews in order to detect inconsistency and avoid random reviews.

To manage the entire review process, we developed many Python scripts working on files exported by the CMT system. The complete set of Python scripts (and related Jupyter notebooks) used to handle the whole review process is also available online\footnote{Accessible at: https://github.com/mscarpiniti/IJCNN2025}.

\section{Conclusion}
\label{sec:conclusion}

In this paper, we have shown the procedures implemented to handle the whole review process of the 2025 International Joint Conference on Neural Networks (IJCNN 2025) along with the main challenges encountered during the process. The choices behind the Reviewers selection and assignment have been analyzed, while the procedure implemented to obtain a ranking of the submitted papers by means of a proposed score index has been discussed. In addition, an experimental procedure to remove reviewer-specific biases has also been discussed.

We hope that this paper and the discussed material could help the Chairs of the next editions of this premier international conference in the area of neural networks and that the hints here outlined could simplify their future work.

\section*{Acknowledgment}
\label{sec:ack}

We would like to express our deepest gratitude to all who contributed to the success of IJCNN 2025: the whole Organizing Committee, the International Neural Network Society (INNS) Board of Governors, the Conference Sponsors. We are also grateful to Area Chairs and Reviewers, the kind and valuable staff of Conference Catalyst and Symposia, and the countless Volunteers and staff behind the scene. A special thank you to the authors and attendees who continue to bring their best work and curiosity to our community.


\bibliographystyle{IEEEtran}
\bibliography{IJCNN2025}

\end{document}